\documentclass[%
showpacs,preprintnumbers,
 amsmath,amssymb,
 aps,
 prd,
 longbibliography,
floatfix,
 lengthcheck,%
]{revtex4-1}

\usepackage{graphicx}
\usepackage{hyperref}
\usepackage{epstopdf}
\usepackage{bbold}

\newcommand{\ket}[1]{\ensuremath{|{#1}\rangle}}

\begin{document}

\title{Multi-flavor effects in Stimulated Transitions of Neutrinos}

\author{Y.\ Yang}
\affiliation{Department of Physics, North Carolina State University,Raleigh,North Carolina 27695, USA}

\author{J.\ P.\ Kneller}
\affiliation{Department of Physics, North Carolina State University,Raleigh,North Carolina 27695, USA}

\author{K.\ M.\ Perkins}
\affiliation{Department of Physics, North Carolina State University,Raleigh,North Carolina 27695, USA}

\begin{abstract}
A neutrino subject to an external, time-dependent perturbing potential can be forced to make transitions between its flavor states. A neutrino with three (or more) flavors can exhibit phenomena that cannot occur if the neutrino had just two. We present an approximate analytic solution for the temporal evolution of a multi-flavor neutrino in response to an arbitrary perturbing Hamiltonian that has been decomposed into its Fourier modes. We impose no restriction upon the number of flavors nor upon the structure of the perturbing Hamiltonian, the number of Fourier modes, their amplitude or their frequencies. We apply the theory to study three-flavor neutrino transformation due to perturbations built from two and three Fourier modes. For the case of two Fourier modes we observe the equivalent of ``induced transparency'' from quantum optics whereby transitions between a given pair of states are suppressed due to the presence of a resonant mode between another pair. When we add a third Fourier mode we find a new effect whereby the third mode can manipulate the transition probabilities of the two mode case so as to force complete transparency or, alternatively, restore ``opacity'' meaning the perturbative potential regains its ability to induce neutrino transitions. In both applications we demonstrate how the analytic solutions are able to match the amplitude and wavenumber of the numerical results to within a few percent. 
\end{abstract}

\maketitle

%
%

\section{Introduction}

Determining the response of a quantum mechanical system to a time-dependent perturbation is a frequent endeavor of both experiment and theory in a large number of subfields of physics. A number of phenomena have been found to occur in quantum optics, in electronic spin and nuclear magnetic resonance, and in ultracold atoms and molecules to name just a few. Many reviews of driven quantum systems can be found including \cite{2015PhyS...90h8013C}. From the theory side, several approaches to the calculation of the transition probability between the states of the system can be found in textbook literature with the various techniques having strengths and weaknesses depending upon the form of the perturbation. Even when we restrict our attention to harmonic perturbations, one may compute the transition probability (or transition rate) between states using Floquet theory \cite{1965PhRv..138..979S}, the Rotating Wave Approximation (RWA) \cite{1940PhRv...57..522B}, Fermi's Golden Rule and others. A comparison between these techniques for a two-level system can be found in Dion \& Hirschfelder \cite{1976acp....35..265D}. 

Perhaps one of the more unusual cases of a driven quantum mechanical system is for the flavor evolution in space/time of a neutrino as it propagates through matter with periodic fluctuations. Due to the difference in neutrino masses, a $N_f$ flavor neutrino has $N_f$ distinct eigenstates. Both Floquet theory and the Rotating Wave Approximation approaches have been used to calculate the effect of density fluctuations described by a single Fourier mode (FM) upon a two flavor neutrino \cite{Ermilova,Akhmedov,1989PhLB..226..341K,1999NuPhB.538...25A,2001PAN....64..787A,Kneller:2012id}. More recently Patton, Kneller and McLaughlin \cite{2014PhRvD..89g3022P} considered the case of two flavor neutrino evolution through \emph{aperiodic} matter fluctuations as one would find in a turbulent medium. Patton, Kneller and McLaughlin based their theoretical description of the evolution upon the Rotating Wave Approximation (RWA) and found it gave predictions which were in remarkably good agreement with numerical calculations on a case -by-case basis. They named their model Stimulated Transitions. 
The evolution of neutrinos through random matter density fluctuations has been previously considered by many others \cite{1987PhLB..185..417S,1990PhRvD..42.3908S,Loreti:1995ae,Fogli:2006xy,Friedland:2006ta,2010PhRvD..82l3004K,Reid,2013PhRvD..88b3008L,2003PhRvD..68c3005F,Choubey,1996NuPhB.472..495N,1998PhRvD..57.3140H,2009PhLB..675...69K,1996PhRvD..54.3941B,Burgess} using statistical approaches. Note an alternative analysis of the similar problem of a two-level atom interacting with a stochastic electromagnetic field is found in Cummings \cite{1982NCimB..70..102C}. 

But the two level quantum system / two flavor neutrino is a restricted case. It is well known that one finds richer phenomenology when the system possesses three or more eigenstates. Perhaps the best known examples are in the field of quantum optics where one has the phenomenon of electromagnetic induced transparency \cite{1990PhRvL..64.1107H,1991PhRvL..66.2593B,2003RvMP...75..457L,2005RvMP...77..633F} and coherent population trapping into dark states \cite{2012Sci...338.1609D} which has also been seen in quantum dots and solid-state systems  \cite{2008NatPh...4..692X,2011Natur.478..497T}. Since neutrinos have (at least) three flavors, not two, one wonders if some of the phenomenology of the three eigenstate systems can be found for a three flavor neutrino. 

The goal of this paper is to present the generalization of the Stimulated Transition model developed by Patton, Kneller and McLaughlin to the case of multi-flavor neutrinos and subject to an arbitrary, Fourier-decomposed perturbing Hamiltonian. In order to facilitate applications to other quantum systems, the theory is described in section \S\ref{sec:theory} in the most general terms possible without reference to any particular quantum system. We then apply the theory in section \S\ref{induced} to three-flavor neutrinos passing through density fluctuations composed of two and three anharmonic FMs. We find the neutrino equivalent of electromagnetic induced transparency for two FMs, and then a new effect we call Restored Opacity when we consider three FMs. Our conclusions are presented in section \S\ref{conclusions}.

%
%

\section{Stimulated Transformation}
\label{sec:theory}

We begin with the general problem of the evolution in time of a arbitrary $N$ level quantum system due to a time dependent perturbation. At some initial time $t_1$ we prepare the system in some arbitrary state - represented by a column vector - which we decompose in terms of the $N$ eigenstates of some basis $(X)$. The system then evolves to a time $t_2$ and at which we decompose the state in terms of the $N$ eigenstates of a possibly different basis $(Y)$. The evolution is described by a matrix $S^{(YX)}(t_2,t_1)$ and the transition probabilities are the set of probabilities that the system in a given initial state $x$ of $(X)$ at $t_1$ is detected in the state $y$ of $(Y)$ at $t_2$. These transition probabilities are denoted by $P^{(YX)}_{yx}(t_2,t_1)$ and are related to the elements of $S^{(YX)}$ by $P^{(YX)}_{yx} = |S^{(YX)}_{yx}|^2$. Since $S^{(YX)}$ must be unitary, one needs $N^2$ independent real parameters in order to describe the matrix $S^{(YX)}$ but note only $(N-1)^2$ of the elements of $P^{(YX)}$ are independent. Hereafter we shall work with the case where the bases $(X)$ and $(Y)$ are the same although there are certainly circumstances where knowing the evolution from one basis to a different basis is useful. Note also that throughout this paper we set $\hbar=c=1$.

In the generic basis $(X)$ the evolution matrix can be found by solving the Schr\"{o}dinger equation 
\begin{equation}
\imath {\frac{dS}{dt}}^{(XX)} = H^{(X)}\,S^{(XX)} \label{eq:dSdt}
\end{equation}
where $H^{(X)}$ is the Hamiltonian in the basis $(X)$. The initial condition is $S^{(XX)}(t_1,t_1) = 1$. We make no assumption about the structure of $H^{(X)}$ except that it be possible to separate the Hamiltonian into an unperturbed piece $\breve{H}^{(X)}(t)$ and a position dependent perturbation $\delta H^{(X)}(t)$ i.e.\ $H^{(X)}(t) = \breve{H}^{(X)}(t) + \delta H^{(X)}(t)$.

If $\breve{H}^{(X)}(t)$ is not diagonal then we introduce an instantaneous unperturbed eigenbasis $(u)$ by finding the unitary matrix $\breve{U}(t)$ defined by $\breve{H}^{(X)} = \breve{U}\,\breve{K}\breve{U}^{\dagger}$ where $\breve{K}$ is the diagonal matrix of the eigenvalues of $\breve{H}$, that is $\breve{K} = \rm{diag}( \breve{k}_1,\breve{k}_2,\ldots)$. The evolution matrix in the instantaneous unperturbed eigenbasis is related to the evolution $S^{(XX)}$ by $S^{(uu)}(t_2,t_1) = \breve{U}^{\dagger}(t_2) S^{(XX)}(t_2,t_1)\breve{U}(t_1)$ 
In this unperturbed eigenbasis
\begin{equation}
H^{(u)} = \breve{K} - \imath \breve{U}^{\dagger}\,\frac{d\breve{U}}{dt} + \breve{U}^{\dagger}\delta H^{(X)}\breve{U} 
\end{equation} 
We now write the evolution matrix in the unperturbed eigenbasis as the product $S^{(uu)} = \breve{S}\,A$ where $\breve{S}$ is defined to be the solution of 
\begin{equation}
\imath \frac{d\breve{S}}{dt} = \left[ \breve{K} - \imath \breve{U}^{\dagger}\,\frac{d\breve{U}}{dt} \right] \,\breve{S}.
\end{equation}
If we know the solution to the unperturbed problem, $\breve{S}$, we can solve for the effect of the perturbation by finding the solution to the differential equation for $A$: 
\begin{equation}
\imath \frac{dA}{dt} = \breve{S}^{\dagger}\,\breve{U}^{\dagger}\delta H^{(X)}\breve{U}\,\breve{S} \,A. \label{dAdt}
\end{equation}
In general the term $\breve{U}^{\dagger}\delta H^{(X)}\breve{U}$ which appears in this equation possesses both diagonal and off-diagonal elements. The diagonal elements are easily removed by writing the matrix $A$ as $A=W\,B$ where $W=\exp(-\imath\Xi)$ and $\Xi$ a diagonal matrix $\Xi=\rm{diag}(\xi_{1},\xi_{2},\ldots)$. Substitution into (\ref{dAdt}) gives a differential equation for $B$
\begin{equation}\label{dBdt}
\imath \frac{dB}{dt} = W^{\dagger}\left[\breve{S}^{\dagger}\breve{U}^{\dagger}\delta H^{(X)}\breve{U}\,\breve{S} -\frac{d\,\Xi}{dt}\right]\,W\,B \equiv H^{(B)} B
\end{equation}
and $\Xi$ is chosen so that $d\,\Xi/dt$ removes the diagonal elements of $\breve{S}^{\dagger}\breve{U}^{\dagger}\delta H^{(X)}\breve{U}\,\breve{S}$. Once $\Xi$ has been found, determining transition probabilities is reduced to solving for the $B$ matrix.

%
%

\subsection{Fourier-decomposed Perturbations}

We now consider the specific case of a constant potential for $\breve{H}^{(X)}$. This form for $\breve{H}$ means $\breve{S}$ is a diagonal matrix $\breve{S}=\exp(-\imath \breve{K}\,t)$. The perturbation $\delta H$ is taken to be a Fourier-like series of the form 
\begin{equation}\label{HM} 
 \delta H^{(X)} = \sum_a (C_{a} e^{\imath q_{a} t} + C_{a}^{\dagger} e^{-\imath q_{a} t})
\end{equation}
where $C_a$ is an arbitrary complex matrix and $q_a$ the frequency of the $a^{th}$ FM of the perturbation. We make no restriction on the number of FMs, the frequencies $q_a$ nor the size or structure of the matrices $C_a$. This generalization to arbitrary structure for the $C_a$'s is where we depart from previous analyses by Patton, Kneller \& McLaughlin \cite{2014PhRvD..89g3022P}. We also refer the reader to Brown, Meath \& Tran \cite{Brown:Meath:Tran} and Avetissian, Avchyan \& Mkrtchian \cite{2012JPhB...45b5402A} who considered the related but simpler problem of the effect of two lasers of different colors, i.e.\ two FMs, upon a two-level dipolar molecule.

Given this form for the perturbation, equation (\ref{dBdt}) indicates we need to consider the combination $\breve{U}^{\dagger}C_{a}\breve{U}$. If we write the diagonal elements of $\breve{U}^{\dagger}C_{a}\breve{U}$ as 
\begin{equation}
 {\rm diag}( \breve{U}^{\dagger}C_a\breve{U} ) = \frac{F_a}{2\imath}  \exp(\imath\,\Phi_a)
 \label{eq7}
\end{equation}
where $F_a$ is a diagonal matrix of amplitudes $F_a={\rm diag}(f_{a;1},f_{a;2},\ldots)$ and $\Phi_a$ the diagonal matrix of phases $\Phi_a={\rm diag}(\phi_{a;1},\phi_{a;2},\ldots)$, then the matrix $\Xi$ is found to be $\Xi(t) = \sum_a \Xi_a(t)$
with 
\begin{equation}
\Xi_a(t) = \frac{F_a}{q_a} \big[ \cos\Phi_a - \cos(\Phi_a + q_a\,t) \big].
\end{equation} 
We denote the diagonal elements of $\Xi_a$ as $\xi_{a;1}, \xi_{a;2},\ldots$. Next we rewrite the off-diagonal elements of $\breve{U}^{\dagger}C_a\breve{U}$ as a matrix $G_a$ i.e\ $G_a ={\rm offdiag}(\breve{U}^{\dagger}C_a\breve{U})$. Putting together the solution for $\Xi$ and $\breve{S}$ and inserting the new matrix $G_a$, we find the Hamiltonian for $B$ is 
\begin{equation}
\begin{split}
 H^{(B)} =&\exp(\imath\,\Xi)\exp(\imath\,\breve{K}\,t)\,\Big(\sum_a \left[ G_a e^{\imath q_a\,t} + G^{\dagger}_a e^{-\imath q_a\,t}  \right] \Big)\, \\
& \times \exp(-\imath\,\breve{K}\,t) \exp(-\imath\,\Xi) 
\end{split}
\end{equation} 
Written explicitly the element $ij$ of the Hamiltonian is
\begin{equation}
 H_{ij}^{(B)} = e^{\imath (\delta\breve{k}_{ij}t+\delta\xi_{ij})} \sum_a\Big[ G_{a;ij}e^{\imath q_a\,t} + G^{\star}_{a;ji}e^{-\imath q_a\,t}\Big]
\end{equation} 
where $\delta\breve{k}_{ij} = \breve{k}_{i}-\breve{k}_{j}$ and $\delta\xi_{ij} = \xi_{i}-\xi_{j}$. 
\\
\\
If we define 
\begin{eqnarray}
x_{a;ij} & = & \frac{f_{a;i}}{q_a}\cos\phi_{a;i} - \frac{f_{a;j}}{q_a}\cos\phi_{a;j} \label{eq11}\\
y_{a;ij} & = & \frac{f_{a;i}}{q_a}\sin\phi_{a;i} - \frac{f_{a;j}}{q_a}\sin\phi_{a;j}  \label{eq12}\\
z_{a;ij} & = & \sqrt{x_{a;ij}^{2}+y_{a;ij}^{2}}   \label{eq13}\\
\psi_{a;ij} & = & \arctan\left( {\frac{{{y_{a;ij}}}}{{{x_{a;ij}}}}} \right) \label{eq14}
\end{eqnarray}

then the term $\delta\xi_{ij}$ is equal to
\begin{equation}
 \delta\xi_{ij} = \sum_a\Big[ x_{a;ij} - z_{a;ij}\cos(q_a\,t+\psi_{a;ij} )\Big].
\end{equation}
The presence of $y_{a;ij}$ in these equations is a new feature of the more general perturbing Hamiltonian we are considering. We now make use of the Jacobi-Anger expansion for $e^{\imath\delta\xi_{ij}}$ 
\begin{widetext}
\begin{equation}
e^{\imath\delta\xi_{ij}} = \prod_a \left\{ e^{\imath x_{a;ij}} \sum_{m_a=-\infty}^{\infty} (-\imath)^{m_a} J_{m_a}(z_{a;ij})\exp\Big[\imath\,m_a\left(q_a\,t + \psi_{a;ij}\right)\Big] \right\}.
\end{equation}
If we substitute this expansion into the expression for the elements of $H^{(B)}$ and define $\mu_{a,m_a;ij}$ and $\lambda_{a,m_a;ij}$ to be
\begin{eqnarray}
\lambda_{a,m_a;ij} & = & (-\imath)^{m_a}\,e^{\imath x_{a;ij}}\,J_{m_a}(z_{a;ij})\,e^{\imath m_a\psi_{a;ij}}
\label{eq17}\\
\mu_{a,m_a;ij} & = & (-\imath)^{m_a}\,e^{\imath x_{a;ij}}\,\left[ G^{\star}_{a;ji}\,J_{m_a+1}(z_{a;ij})\,e^{\imath(m_a+1)\psi_{a;ij}}-G_{a;ij}\,J_{m_a-1}(z_{a;ij})\,e^{\imath(m_a-1)\psi_{a;ij}}\right] 
\end{eqnarray}
then we find the element $ij$ of the Hamiltonian is given by 
\begin{equation}\label{HB_ij}
 H_{ij}^{(B)} = \imath \sum_a\left\{ \sum_{m_a}\mu_{a,m_a;ij}\,e^{\imath ( m_a q_a + \delta\breve{k}_{ij}) t}\prod_{b\neq a}\left[ \sum_{m_b} \lambda_{b,m_b;ij}\,e^{\imath m_b q_b t} \right]. \right\}
\end{equation} 
\end{widetext}


\subsection{Rotating Wave Approximation}

Even though we started with a very general perturbing Hamiltonian, we have found a form for $H^{(B)}$ which has the same structure as that found by Patton, Kneller \& McLaughlin. From here on, we follow the same procedure to solve for the matrix $B$. First we adopt the Rotating Wave Approximation. The RWA amounts to selecting a particular value for the integers $m_a$ and $m_b$ in Eq.(\ref{HB_ij}) and dropping all others.
We do not specify a procedure for selecting those integers though algorithms exist. We expect there is not one procedure that can be adopted universally for all situations. There are some restrictions to be placed on the selection of the integers. In order that the resulting Hamiltonian be solvable we cannot make choices for $m_a$ and $m_b$ for every element $ij$ independently. Only $N-1$ elements are to be regarded as independent and a suitable set could be either those on the sub/superdiagonal or the off-diagonal elements in a particular row or column. The values of  $m_a$ and $m_b$ we select will be different for each independent element. We denote these integers by $n_{a;ij}$ since they are specific both to the frequency $a$ and the element of the Hamiltonian $ij$, and define 
\begin{equation}
 \kappa_{ij} = \sum_a\mu_{a,n_{a;ij};ij} \prod_{b\neq a} \lambda_{b,n_{b;ij};ij} 
\end{equation}
then $H_{ij}^{(B)}$ is simplified to
\begin{equation}
H_{ij}^{(B)} = -\imath \kappa_{ij} \exp\left[\imath \left( \sum_a n_{a;ij}\, q_a + \delta\breve{k}_{ij}\right) t\right],
\end{equation} 
or, in full matrix, form 
\begin{widetext}
\begin{equation}
H^{(B)} = \left( \begin{array}{cccc}
0 & - \imath \kappa_{12}\,e^{\imath \left[ \delta\breve{k}_{12} + \sum\limits_a n_{a;12}\,q_a \right]t} & - \imath \kappa_{13}\,e^{\imath \left[ \delta\breve{k}_{13} + \sum\limits_a n_{a;13}\,q_a \right]t} & \ldots \\
\imath \kappa^{\star}_{12}\,e^{-\imath \left[ \delta\breve{k}_{12} + \sum\limits_a n_{a;12}\,q_a \right]t} & 0 & \imath \kappa_{12}\,e^{-\imath \left[ \delta\breve{k}_{23} + \sum\limits_a n_{a;23}\,q_a \right]t} & \ldots \\
\imath \kappa^{\star}_{23}\,e^{-\imath \left[ \delta\breve{k}_{13} + \sum\limits_a n_{a;13}\,q_a \right]t} & \imath \kappa^{\star}_{23}\,e^{-\imath \left[ \delta\breve{k}_{23} + \sum\limits_a n_{a;23}\,q_a \right]t} & \ldots \\
 \vdots & \vdots & \vdots & \ddots 
\end{array} \right)
\label{eq:hb}
\end{equation}
\end{widetext}
Again, we remind the reader that, for example, only $n_{a;12}$, $n_{a;23}$, $n_{a;34}$ etc.\ are independent: in all other cases the integer $n_{a;ij} = n_{a;i\ell} + n_{a;\ell j}$. As shown by Patton, Kneller \& McLaughlin, with this simplified Hamiltonian, Eq.(\ref{dBdt}), can be solved for the evolution matrix $B$ and we reproduce their solution here for completeness. Since both $n_{a;ij} = n_{a;i\ell} + n_{a;\ell j}$ and $\delta\breve{k}_{ij} = \delta\breve{k}_{i\ell} + \delta\breve{k}_{\ell j}$, we can factorize $H^{(B)}(t)$ into the form $H^{(B)}(t) = \Upsilon(t)\,M\,\Upsilon^{\dagger}(t)$ where the matrix $M$ is a constant, i.e. it contains the couplings $\kappa_{ij}$ only. The matrix $\Upsilon$ is of the form $\Upsilon(t)=\exp(\imath\,\Lambda\,t)$, where $\Lambda$ is also a constant matrix that depends only on $\delta\breve{k}_{ij}$, the integer sets $\{n_{a;ij}\}$ and 
the frequencies $q_a$. Explicitly we can write 
\begin{equation}\label{eq for M}
M = \left( \begin{array}{cccc}
    0 & -\imath \kappa_{12} & -\imath \kappa_{13} & \ldots \\
    \imath \kappa^{\star}_{12} & 0 & -\imath \kappa_{23} & \ldots \\
    \imath \kappa^{\star}_{13} & \imath \kappa^{\star}_{23} & 0 & \ldots \\
    \vdots & \vdots & \vdots & \ddots
    \end{array}\right),
\end{equation}
and one possible choice for the matrix $\Lambda$ is 
\begin{widetext}
\begin{equation}
\Lambda  = \left( {\begin{array}{cccc}
\breve{k}_{1} + \sum\limits_a n_{a;1}\,q_a & 0 & 0 & \ldots \\
0 & \breve{k}_{2} + \sum\limits_a n_{a;2}\,q_a & 0 & \ldots \\
0 & 0 & \breve{k}_{3} + \sum\limits_a n_{a;3}\,q_a & \ldots \\
 \vdots & \vdots & \vdots & \ddots 
\end{array}} \right),
\end{equation}
\end{widetext}
where $n_{a;i}$ are integers chosen so that $n_{a;i} - n_{a;j} = n_{a;ij}$. Using this factorization of $H^{(B)}(t)$ we find equation (\ref{dBdt}) can be rewritten as   
\begin{equation}
\imath\Upsilon^{\dagger}\frac{dB}{dt} =  M\Upsilon^{\dagger}\,B
\end{equation}
Instead of solving for $B$ we solve for the combination $\Omega=\Upsilon^{\dagger} B$. The differential equation for $\Omega$ is found to be 
\begin{equation}
\imath\frac{d\Omega}{dt} = \left(M +\Lambda \right)\,\Omega = H^{(\Omega)}\,\Omega.
\end{equation}
Since the matrix both $M$ and $\Lambda$ are constant matrices, the matrix $H^{(\Omega)}$ is also independent of $t$ meaning $\Omega$ has the formal solution $\Omega(t) = \exp(-\imath H^{(\Omega)} t)\,\Omega(0)$. The solution for $B$ is thus
\begin{equation}
B(t) = \Upsilon(t)\,\exp(-\imath H^{(\Omega)} t)\,\Upsilon^{\dagger}(0) B(0). \label{eq:soln for B}
\end{equation}
Now that we have the solution for $B$, the full evolution matrix in the basis $(u)$ is $S=\breve{S}\,W\,B$ but given that both $\breve{S}$ and $W$ are diagonal matrices, the transition probability between the unperturbed eigenstates is simply the square magnitude of the off-diagonal element of $B$. Note that we have made no restrictions on the number of FMs, the frequencies nor the size and structure of the amplitude matrices so this procedure for obtaining the RWA solution applies to both periodic \emph{and} aperiodic Hamiltonians. 

For the particular case of two eigenstates we can write out the solution succinctly (after dropping the $12$ subscripts on $n_{a;12}$ and $\kappa_{12}$) by introducing the detuning frequency $p$ via $2p = \delta\breve{k}_{12} + \sum\limits_a {n_a q_a}$ and the Rabi flopping frequency $Q$ by $Q^{2} = p^{2}+ \kappa^{2}$. Using these quantities, $B(t)$ for two flavors is found to be 
\begin{widetext}
\begin{equation}
B = \left( \begin{array}{cc}
 e^{\imath p t} \left[ \cos\left(Q\,t\right) -\imath\frac{p}{Q}\sin\left(Q\,t\right) \right] & -e^{\imath p t}\,\frac{\kappa}{Q}\sin\left(Q\,t\right) \\
e^{-\imath p t}\frac{\kappa^{\star}}{Q}\sin\left(Q\,t\right) & e^{-\imath p t} \left[ \cos\left(Q\,t\right) +\imath\frac{p}{Q}\sin\left(Q\,t\right) \right]. 
 \end{array} \right)
\end{equation}
\end{widetext}
and so we see the transition probability between the matter states 1 and 2 is 
\begin{equation}
P_{12} = |B_{12}|^{2}= \frac{\kappa^2}{Q^2}\,\sin^{2}\left(Q\,t\right). 
\label{transition}
\end{equation}
This is the result found by Patton, Kneller and McLaughlin \cite{2014PhRvD..89g3022P}. A transition probability that varies as $\sin^2 (Q\,t)$ is a generic prediction of \emph{all} RWA solutions of a two-level quantum system when the perturbing Hamiltonian can be expressed as a Fourier-like series even if the ratios between the frequencies of different modes are not rational numbers, which makes the series aperiodic.


\subsection{Degenerate RWA}
If the FMs are such that the ratio between any pair of frequencies, $q_a$ and $q_b$, is a rational fraction then there are multiple sets of the integers $\{n\}$ which \emph{all} have the same detuning frequency. One can account for these degenerate rotating waves by simply adding them together so that $\kappa_{ij}$ becomes

\begin{equation}
\kappa_{ij} = \sum_{\{n\}} \left[ \sum_a\mu_{ij,n_{a;ij}} \prod_{b\neq a} \lambda_{ij,n_{b;ij}} \right].
\end{equation}
If all ratios between the frequencies are rational then accounting for the degeneracy in the RWA becomes an exercise in combinatorics. Even so, typically one finds the sum is dominated by one set - the one where the sum of the absolute values of the integers $\{n\}$ is smallest. 


\section{Application to Neutrino Propagation}
\label{induced}

With the theory complete we now move on to testing whether it gives predictions which agree with numerical solutions. The problem we choose is the case of a three-flavor neutrino with an energy of $5\;{\rm MeV}$ propagating through fluctuating matter. Note, as commonly found in the literature on neutrino flavor transformation, we switch the variable from time $t$ to position along the neutrino trajectory $r$ where $r=c\,t$ since the neutrino wavepacket is localized in space and typically the energy of neutrinos is much larger than their rest mass hence they move at a speed close to $c$. As a consequence we switch from using `frequencies' to `wavenumbers'.  

The Hamiltonian, $H$, governing the neutrino flavor evolution through matter is the sum of a constant vacuum term $H_V$ and a term 
coming from the effect of matter $H_{M}$ \cite{Wolfenstein1977,M&S1986}. The vacuum Hamiltonian in the `flavor basis' is 
\begin{equation}
H^{(f)}_V = \frac{1}{2E} U_V \left( \begin{array}{*{20}{c}} m_1^2-m_2^2 & 0 & 0 \\ 0 & 0 & 0 \\ 0 & 0 & m_3^2-m_2^2 \end{array} \right) U_V^{\dagger}
\end{equation}
\\
where $U_V$ is the vacuum mixing matrix and $m_i$ the three neutrino masses. We set the squared mass differences $m_1^2-m_2^2 = -7.5 \times 10^{-5}\;{\rm eV^2}$ and $m_3^2-m_2^2 =2.32\times 10^{-3}\;{\rm eV^2}$ which are compatible with the mass-squared differences as given by the Particle Data Group \cite{PDG}. $U_V$ is parameterized by three mixing angles $\theta_{12}$, $\theta_{13}$ and $\theta_{23}$ - we set all possible phases to zero \cite{2009PhRvD..80e3002K} - and given by 
\begin{widetext}
\begin{eqnarray}
U & = & 
\left(\begin{array}{lll} c_{12}c_{13} & s_{12}c_{13} & s_{13}
\\ 
-s_{12}c_{23}-c_{12}s_{13}s_{23}
& c_{12}c_{23}-s_{12}s_{13}s_{23}
& c_{13}s_{23} \\ 
s_{12}s_{23}-c_{12}s_{13}c_{23} &
-c_{12}s_{23}-s_{12}s_{13}c_{23}
& c_{13}c_{23} \end{array}\right) 
\label{eq:U}
\end{eqnarray}
\end{widetext}
where the notation is that $c_{ij} = \cos\theta_{ij}$ and $s_{ij} = \sin\theta_{ij}$. We take the angles to be $\theta_{12} =34^{\circ}$, $\theta_{13} =9^{\circ}$ and $\theta_{23} =45^{\circ}$ \cite{PDG}. 

\subsection{Two Fourier Modes}

We first consider the case where the matter Hamiltonian is taken to be a constant upon which are superposed two FMs with wavenumbers $q_1$ and $q_2$ which are not in a rational ratio. The matter is regarded as affecting only the electron flavor type, not the other two flavors. The form of the Hamiltonian in the flavor basis with the first row/column indicating the electron flavor is thus 
\begin{equation}
H^{(f)}_M(r) = V_{\star} \left[1 + A_1 \cos(q_1 r) + A_2 \cos(q_2 r)\right] \left( \begin{array}{lll} 1 & 0 & 0 \\ 0 & 0 & 0 \\ 0 & 0 & 0 \end{array} \right)
\end{equation}
with $V_{\star}$ the potential from the constant background, $A_1$ and $A_2$ the amplitudes of the fluctuations. In what follows we set $V_{\star}$ to $V_{\star}=6\times 10^{-25}\;{\rm erg}$.

\begin{figure}[t]
\includegraphics[clip,width=\linewidth]{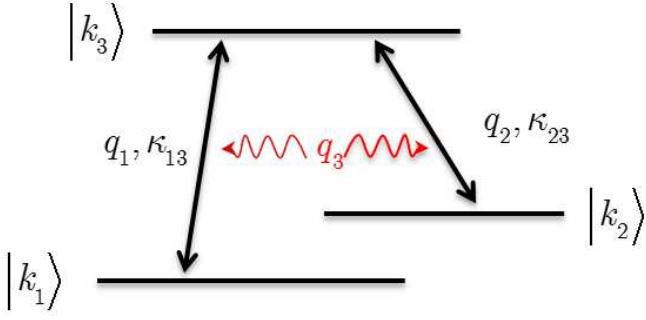}
\caption{The three unperturbed eigenstates and the transitions between them in the two and three FM problems. The modes $q_1$ and $q_2$ are the FMs that drive transitions between the indicates states. The mode $q_3$ is the ``switch mode'' which switches on and off the effect of transitions induced by modes $q_1$ and $q_2$.}
\label{fig:levels}
\end{figure}

The vacuum Hamiltonian and the constant potential $V_{\star}$ form the `unperturbed' Hamiltonian $\breve{H}$. In the flavor basis $\breve{H}$ is not diagonal. We can diagonalize $\breve{H}$ by first finding its eigenvalues, denoted by $\breve{K} ={\rm diag}\left( \breve{k}_1, \breve{k}_2, \breve{k}_3\right)$, and then the unitary matrix $\breve{U}$ which satisfies $\breve{H}^{f} = \breve{U}\,\breve{K}\,{\breve{U}}^{\dagger}$. Since this is standard textbook quantum mechanics, we leave this as an exercise for the reader. For reference, the differences between the eigenvalues are found to be $\breve{k}_3 - \breve{k}_1 = 3.835\times 10^{-22}\;{\rm erg}$ and $\breve{k}_3 - \breve{k}_2 = 3.715\times 10^{-22}\;{\rm erg}$. Note that since $\breve{H}$ is a function of $V_{\star}$, the eigenvalues and unperturbed mixing matrix, $\breve{U}$ are also functions of $V_{\star}$. The level scheme we end up with is shown in figure (\ref{fig:levels}) with the three eigenstates of the unperturbed system denoted as 
$\ket{k_1}$, $\ket{k_2}$ and $\ket{k_3}$.

The two FMs in the matter Hamiltonian are the Fourier-decomposed perturbation. The $C$ matrices for the two FMs are 
\begin{equation}
 C_1 = \frac{1}{2} V_{\star} A_1 \left( \begin{array}{lll} 1 & 0 & 0 \\ 0 & 0 & 0 \\ 0 & 0 & 0 \end{array} \right) \label{eq:C1}
\end{equation}
and
\begin{equation}
 C_2 = \frac{1}{2} V_{\star} A_2 \left( \begin{array}{lll} 1 & 0 & 0 \\ 0 & 0 & 0 \\ 0 & 0 & 0 \end{array} \right) \label{eq:C2}
\end{equation}
giving diagonal elements of the $F_a$ matrices of the form 
\begin{equation}
f_{a;i} = \frac{1}{2}|\breve{U}_{e i}|^{2} A_a V_{\star} \label{eq:f_a;i}
\end{equation}
while all the elements in the $\Phi_a$ matrices are $\pi/2$. This has the consequence that for both FMs 
\begin{equation}
y_{a;ij} = z_{a;ij} =\frac{ A_a V_{\star} \left( |U_{ei}|^2 - |U_{ej}|^2 \right) }{2\,q_a} \label{y_a;ij}
\end{equation}
and therefore all of the phases $\psi_{a;ij}$ are also equal to $\pi/2$. The elements of the $G_a$ matrices are 
\begin{equation}
G_{a;ij} = \frac{1}{2}\,\breve{U}_{e i}^{\star}\breve{U}_{e j}\,A_a V_{\star} \label{G_a;ij}
\end{equation}
If we put all these pieces together we find the $\lambda$'s are simply
\begin{equation}
\lambda_{a,m_a;ij} = J_{m_a}\left(z_{a;ij} \right) \label{lambda_a,ma;ij}
\end{equation}
while the $\kappa$'s are of the form 
\begin{equation}
 \kappa_{a,m_a;ij} = \imath\left[ {{G_{a;ji}}{J_{{m_a} + 1}}\left( z_{a;ij} \right) + {G_{a;ij}}{J_{{m_a} - 1}}\left( z_{a;ij} \right)} \right]. \label{kappa_a,ma;ij}
\end{equation}

\begin{figure}[b]
\includegraphics[clip,width=\linewidth]{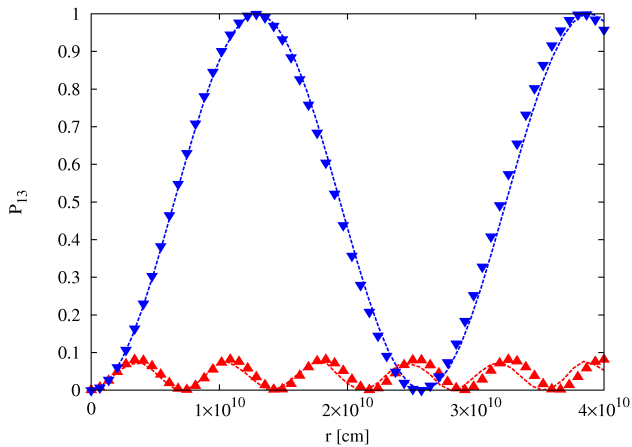}
\caption{The transition probabilities from unperturbed eigenstate 1 to unperturbed eigenstate 3. The blue dashed line is the RWA result for the case $A_1=0.1, A_2=0$ and the red dashed line for $A_1=0.1, A_2=0.5$. The symbols represent the corresponding numerical results.}
\label{fig:ITvsr}
\end{figure}
\begin{figure}[t]
\includegraphics[clip,angle=270,width=\linewidth]{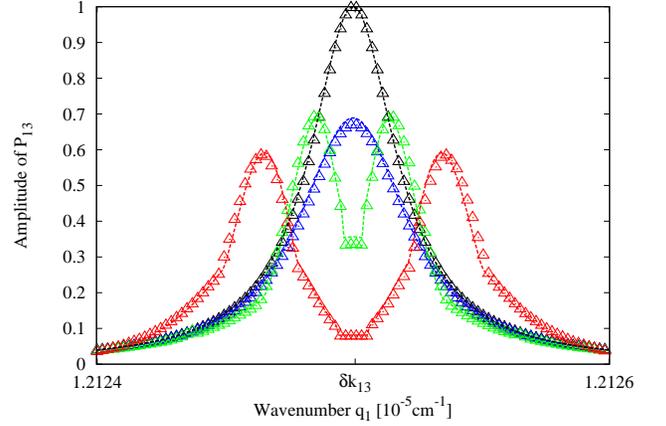}
\caption{The amplitude of $P_{13}$ as a function of $q_1$. The parameters used are $V_{\star}=6\times 10^{-25}\;{\rm erg}, A_1=0.1$, and $A_2=0.5/0.2/0.1/0$ for the black/blue/green/red dashed line and symbols, $A_2=0$ for the blue dashed line and symbols. The symbols represent numerical results, while the dashed lines are from RWA evaluation.}
\label{fig:ITvsq1}
\end{figure}
Let us now set the wavenumbers for the two modes so that $q_1 \approx \breve{k}_3 - \breve{k}_1$ and $q_2 \approx \breve{k}_3 - \breve{k}_2$ as shown in figure (\ref{fig:levels}). The RWA integers we select for the $1,3$ element are thus $\{ n_{1;13}, n_{2;13} \} = \{ +1 , 0 \}$ and for the $2,3$ element we pick $\{ n_{1;23}, n_{2;23} \} = \{ 0 , +1 \}$. The integer set for the $1,2$ element must therefore be $\{ n_{1;12}, n_{2;12} \} = \{ +1 , -1 \}$ in order that $n_{a;12} + n_{a;23} = n_{a;13}$. The Hamiltonian $H^{(B)}$ is thus
\begin{widetext}
\begin{equation}
H^{(B)} = \left( \begin{array}{ccc}
    0 & -\imath \kappa_{12}\, e^{\imath\left(\delta\breve{k}_{12}+q_1-q_2\right) r} & -\imath \kappa_{13}\, e^{\imath\left(\delta\breve{k}_{13} + q_1 \right) r} \\
    \imath \kappa^{\star}_{12}\, e^{-\imath\left(\delta\breve{k}_{12}+q_1-q_2\right) r} & 0 & -\imath \kappa_{23}\, e^{\imath\left(\delta\breve{k}_{23} + q_2\right) r} \\
    \imath \kappa^{\star}_{13}\, e^{-\imath\left(\delta\breve{k}_{13} + q_1\right) r} & \imath \kappa^{\star}_{23}\, e^{-\imath\left(\delta\breve{k}_{23} + q_2\right) r} & 0
    \end{array}\right)
\end{equation}
with
\begin{eqnarray}
\kappa_{12} & = & \mu_{1,n_{1,12};12}\,\lambda_{2,n_{2,12};12} + \lambda_{1,n_{1,12};12}\,\mu_{2,n_{2,12};12} \nonumber \\
     & = &  \imath \,\left[ G_{1;21}\,J_{2}\left(z_{1;12}\right) + G_{1;12}\,J_{0}\left(z_{1;12}\right) \right] \,J_{-1}\left(z_{2;12}\right)
        \;+\; \imath \,J_{1}\left(z_{1;12}\right) \,\left[ G_{2;21}\,J_{0}\left(z_{2;12}\right) + G_{2;12}\,J_{-2}\left(z_{2;12}\right) \right] \\
\kappa_{13} & = & \mu_{1,n_{1,13};13}\,\lambda_{2,n_{2,13};13} + \lambda_{1,n_{1,13};13}\,\mu_{2,n_{2,13};13} \nonumber \\
     & = &  \imath \,\left[ G_{1;31}\,J_{2}\left( z_{1;13} \right) + G_{1;13}\,J_{0}\left(z_{1;13}\right) \right] \,J_{0}\left(z_{2;13}\right)
        \;+\; \imath \,J_{1}\left(z_{1;13}\right) \,\left[ G_{2;31}\,J_{1}\left(z_{2;13}\right) + G_{2;13}\,J_{-1}\left(z_{2;13}\right) \right] \\
\kappa_{23} & = & \mu_{1,n_{1,23};23}\,\lambda_{2,n_{2,23};23} + \lambda_{1,n_{1,23};23}\,\mu_{2,n_{2,23};23} \nonumber \\
     & = &  \imath \,\left[ G_{1;32}\,J_{1}\left(z_{1;23}\right) + G_{1;23}\,J_{-1}\left(z_{1;23}\right) \right] \,J_{1}\left(z_{2;23}\right)
        \;+\; \imath \,J_{0}\left(z_{1;23}\right) \, \left[ G_{2;32}\,J_{2}\left(z_{2;23}\right) + G_{2;23}\,J_{0}\left(z_{2;23}\right) \right] 
\end{eqnarray}
Since we chose the vacuum mixing matrix to be pure real, the unperturbed matter mixing matrix $\breve{U}$ is also pure real. This is a result of the Naumov \cite{1992IJMPD...1..379N} and Harrison \& Scott \cite{2002PhLB..535..229H} identities. In this case we see from equation (\ref{G_a;ij}) that $G_{a;ij} = G_{a;ji}$. We also recall the identities for Bessel functions that $J_{-n}(z)=(-1)^{n}J_{n}(z)$ and $J_{n-1}(z) + J_{n+1}(z) = 2\,n\, J_n(z) / z$ and when combined, these identities mean that the expressions for the $\kappa$'s are the much simpler
\begin{eqnarray}
\kappa_{12} & = & -\frac{2\,\imath \,G_{1;12}}{z_{1;12}} \,J_{1}\left(z_{1;12}\right) \,J_{1}\left(z_{2;12}\right)
             \;+\; \frac{2\,\imath \,G_{2;12}}{z_{2;12}} \,J_{1}\left(z_{1;12}\right) \,J_{1}\left(z_{2;12}\right) \label{eq:kappa12:2}\\
\kappa_{13} & = & \frac{2\,\imath \,G_{1;13}}{z_{1;13}} \,J_{1}\left(z_{1;13}\right) \,J_{0}\left(z_{2;13}\right) \label{eq:kappa13:2}\\
\kappa_{23} & = & \frac{2\,\imath \,G_{2;23}}{z_{2;23}} \,J_{0}\left(z_{1;23}\right) \,J_{1}\left(z_{2;23}\right) \label{eq:kappa23:2}
\end{eqnarray}
\end{widetext}
We notice that both terms in $\kappa_{12}$ are proportional to the product of two Bessel functions $J_1$ so once we recall that the Bessel function $J_{n}(z) \sim z^{|n|}$ for small $z$, we see that the element $\kappa_{12}$ is smaller in magnitude than $\kappa_{13}$ and $\kappa_{23}$ since the values of $z_{a:ij}$ are very smaller. That is confirmed when we compute the numerical values and find $\kappa_{12} = 6.419\times 10^{-32}\,\imath\;{\rm erg}$, $\kappa_{13} = -3.888\times 10^{-27}\,\imath\;{\rm erg}$ and $\kappa_{23} = -1.311\times 10^{-26}\,\imath\;{\rm erg}$. 

We now proceed to solve for $B$. Following the steps given above, the matrix $M$ is 
\begin{equation}
M = \left( \begin{array}{cccc}
    0 & -\imath \kappa_{12} & -\imath \kappa_{13} \\
    \imath \kappa^{\star}_{12} & 0 & -\imath \kappa_{23} \\
    \imath \kappa^{\star}_{13} & \imath \kappa^{\star}_{23} & 0 
    \end{array}\right).
\end{equation}
Next we need the matrix $\Lambda$. In order to construct this matrix we need to find a set of six integers such that $n_{a;i} - n_{a;j} = n_{a;ij}$. The simplest solution is $n_{1;1} = 1$, $n_{1;2} = 0$, $n_{1;3} = 0 $ and $n_{2;1} = 0$, $n_{2;2} = 1$, $n_{2;3} = 0$. With this choice we have

\begin{equation}
\Lambda  = \left( {\begin{array}{ccc}
\breve{k}_{1} + q_{1} & 0 & 0 \\
0 & \breve{k}_{2} + q_{2} & 0 \\
0 & 0 & \breve{k}_{3} 
\end{array}} \right).
\end{equation}
Thus $H^{(\Omega)} = M+\Lambda$ is 
\begin{equation}
H^{(\Omega)} = \left( {\begin{array}{ccc}
\breve{k}_{1} + q_{1} & -\imath \kappa_{12} & -\imath \kappa_{13} \\
\imath \kappa^{\star}_{12} & \breve{k}_{2} + q_{2} &  -\imath \kappa_{23} \\
\imath \kappa^{\star}_{13} & \imath \kappa^{\star}_{23} & \breve{k}_{3}
\end{array}} \right).
\end{equation}
$B$ has the formal solution $B(r) = \Upsilon(r)\,\exp(-\imath H^{(\Omega)} r)\,\Upsilon^{\dagger}(0)$ where $\Upsilon(r)=\exp(\imath\,\Lambda\,r)$ which can be evaluated using standard procedures. If we make the approximation that $\kappa_{12}$ is negligibly small compared to $\kappa_{13}$ and $\kappa_{23}$ and that the two wavenumbers are exactly on resonance, $q_1 = \breve{k}_3 - \breve{k}_1$, $q_2 = \breve{k}_3 - \breve{k}_2$, then we find the analytical expression for the $B$ matrix is
\begin{widetext}
\begin{equation}
B = \exp\left(\imath\,\left[\Lambda-\breve{k}_3\mathbb{1}\right]\,r\right) \,\left( {\begin{array}{*{20}{c}}
{\frac{{{{\left| {{\kappa _{23}}} \right|}^2}}}{{{Q^2}}} + \frac{{{{\left| {{\kappa _{13}}} \right|}^2}}}{{{Q^2}}}\cos \left( {Qr} \right)}&{\frac{{\imath{\kappa _{13}}\kappa _{23}^*}}{{{Q^2}}}\left[ {\cos \left( {Qr} \right) - 1} \right]}&{ - \frac{{{\kappa _{13}}}}{Q}\sin \left( {Qr} \right)}\\
{\frac{{\imath{\kappa _{23}}\kappa _{13}^*}}{{{Q^2}}}\left[ {\cos \left( {Qr} \right) - 1} \right]}&{\frac{{{{\left| {{\kappa _{13}}} \right|}^2}}}{{{Q^2}}} + \frac{{{{\left| {{\kappa _{23}}} \right|}^2}}}{{{Q^2}}}\cos \left( {Qr} \right)}&{ - \frac{{{\kappa _{23}}}}{Q}\sin \left( {Qr} \right)}\\
{\frac{{\kappa _{13}^*}}{Q}\sin \left( {Qr} \right)}&{\frac{{\kappa _{23}^*}}{Q}\sin \left( {Qr} \right)}&{\cos \left( {Qr} \right)}
\end{array}} \right),
\end{equation}
\end{widetext}
where $Q^2 = \left|\kappa_{13}\right|^2 + \left|\kappa_{23}\right|^2$ and $\mathbb{1}$ is a 3x3 unit matrix.
From this result we can extract the transition probability from unperturbed eigenstate 1 to unperturbed eigenstate 3 by taking the squared magnitude of $B_{13}$
\begin{equation}\label{p13}
{P_{13}} = \frac{{{{\left| {{\kappa _{13}}} \right|}^2}}}{{{Q^2}}}{\sin ^2}\left( {Qr} \right) = \left( {1 - \frac{{{{\left| {{\kappa _{23}}} \right|}^2}}}{{{Q^2}}}} \right){\sin ^2}\left( {Qr} \right).
\end{equation}
This result is interesting because it indicates the transition probability $P_{13}$ depends upon the wavenumber $q_2$ which is driving transitions from unperturbed eigenstate 2 to unperturbed eigenstate 3. In the extreme case when $\kappa_{23}$ is significantly larger than $\kappa_{13}$, the transition from states 1 to 3 is strongly suppressed. This is an analog of the Electromagnetically Induced Transparency (EIT) - see, for example, \cite{1990PhRvL..64.1107H,1991PhRvL..66.2593B,2003RvMP...75..457L,2005RvMP...77..633F} - in atomic physics where the presence of a second possible transition between atomic levels 2 and level 3 will inhibit the primary transition from atomic level 1 to level 3 leading to little absorption, and thus transparency, for the light frequency corresponding to the energy splitting of level 1 and 3. 

To illustrate this neutrino version of induced transparency, in Fig.(\ref{fig:ITvsr}) we plot the transition probability as a function of $r$ when the system is at perfect resonance, namely when $q_1 = \breve{k}_3 - \breve{k}_1$ and $q_2 = \breve{k}_3 - \breve{k}_2$. The reader will observe that indeed, even though the wavenumber $q_1$ is exactly on resonance with the transition between neutrino states 1 and 3, the probability of being in state 3 has a maximum of only $10\%$ when $A_2 \neq 0$. When we remove the second FM $q_2$ the transition probability $P_{13}$ increases to $100\%$. Note also a) that the solution is periodic even though the two wavenumbers $q_1$ and $q_2$ do not form rational ratio, and b) how well the numerical solution to the problem agrees with the RWA solution. The predicted amplitude and the wavenumber match the amplitude and wavenumber of the numerical solution to within a few percent. 

To see the effect of induced transparency more clearly, we fix $q_2$ at the resonance between states 2 and 3 and scan in $q_1$. The solution for $B$ can be found by evaluating the formal solution and from the element $B_{13}$ we extract the transition probability $P_{13}$. In Fig.(\ref{fig:ITvsq1}) we plot the amplitude of the oscillations in $P_{13}$ as a function of $q_1$. We see that in the presence of mode $q_2$, the transition probability has a peculiar shape with peaks off-resonance and local minimum at the resonance. If we turn off the second perturbing mode by setting $A_2$ to zero we recover the expected shape for a resonance at $q_1$. Again, we find the RWA is able to reproduce the shape of $P_{13}$ versus $q_1$ very well at all the values of $A_2$ used. 


\subsection{Three Fourier modes}
\label{opacity}

Now we add a third FM to the perturbing Hamiltonian which we give an amplitude $A_3$ and wavenumber $q_3$. Thus the perturbing Hamiltonian in the flavor basis becomes 
\begin{equation}
\delta {H^{\left( f \right)}}\left( r \right) = V_{\star}\sum\limits_{j = 1,2,3} {{A_j}\cos \left( {{q_j}r} \right)} \left( {\begin{array}{*{20}{c}}
1&0&0\\
0&0&0\\
0&0&0
\end{array}} \right)
\end{equation}
We shall leave $V_{\star}$ unchanged so that the unperturbed Hamiltonian is the same as the previous case of two FMs with the same eigenvalues. The structure of the coefficient matrices $C_1$ and $C_2$ are also unchanged from equations (\ref{eq:C1}) and (\ref{eq:C2}) and new coefficient matrix we must introduce for the new mode is structured as
\begin{equation}
 C_3 = \frac{1}{2} V_{\star} A_3 \left( \begin{array}{lll} 1 & 0 & 0 \\ 0 & 0 & 0 \\ 0 & 0 & 0 \end{array} \right)
\end{equation}
Thus the form of $f_{a;i}$, $y_{a;ij}$, $G_{a;ij}$, $\lambda_{a,m_a;ij}$ and $\kappa_{a,m_a;ij}$ are the same as given previous but with the substitution of $q_3$ for the wavenumber. 

Let us now set the wavenumbers for the first two modes so that $q_1 = \breve{k}_3 - \breve{k}_1$ and $q_2 = \breve{k}_3 - \breve{k}_2$ for the level diagram shown in figure (\ref{fig:levels}). Neither $A_1$ nor $A_2$ are zero and $A_2 > A_1$. For a two FM case this choice for the wavenumbers $q_1$ and $q_2$ and ratio of amplitudes would put the system exactly at the midpoint of figure (\ref{fig:ITvsq1}) so the transition probability $P_{13}$ is suppressed even though the wavenumber $q_1$ is exactly on resonance.

We shall not set the mode $q_3$ to a particular value yet but we shall only consider wavenumbers such that $q_3$ is much smaller than $\breve{k}_3 - \breve{k}_1$, $\breve{k}_3 - \breve{k}_2$ and $\breve{k}_2 - \breve{k}_1$. i.e.\ $q_3$ is not on resonance with any pair of eigenvalue splittings. Thus the sets of RWA integers are very similar to the sets for the two FM case: for the $1,3$ element they are $\{ n_{1;13}, n_{2;13}, n_{3;13} \} = \{ +1 , 0, 0 \}$ and for the $2,3$ element we pick $\{ n_{1;23}, n_{2;23}, n_{3;23} \} = \{ 0 , +1, 0\}$. Again, these choices mean the integer set for the $1,2$ element is determined and must therefore be $\{ n_{1;12}, n_{2;12}, n_{3;12} \} = \{ +1 , -1,0 \}$ in order that $n_{a;12} + n_{a;23} = n_{a;13}$. The structure of the Hamiltonian for $H^{(B)}$ is exactly the same as in equation (\ref{eq:hb}).
As before, we find the expressions for the $\kappa$'s simplify greatly once we use the identities given earlier for the case where $G_{a;ij} = G_{a;ji}$, the vacuum mixing matrix is pure real, and from the Bessel functions. The final expressions are:
\begin{widetext}
\begin{eqnarray}
\kappa_{12} & = &  -\frac{2\,\imath\,G_{1;12}}{z_{1;12}} \,J_{1}\left(z_{1;12}\right) \,J_{1}\left(z_{2;12}\right) \,J_{0}\left(z_{3;12}\right)
                    \;+\; \frac{2\,\imath \,G_{2;12}}{z_{2;12}} \,J_{1}\left(z_{1;12}\right) \,J_{1}\left(z_{2;12}\right) \,J_{0}\left(z_{3;12}\right)  \\
\kappa_{13} & = &  \frac{2\,\imath \,G_{1;13}}{z_{1;13}} \,J_{1}\left(z_{1;13}\right) \,J_{0}\left(z_{2;13}\right) \,J_{0}\left(z_{3;13}\right) \\
\kappa_{23} & = &  \frac{2\,\imath \,G_{2;23}}{z_{2;23}} \,J_{0}\left(z_{1;23}\right) \,J_{1}\left(z_{2;23}\right) \,J_{0}\left(z_{3;23}\right)
\end{eqnarray}
\end{widetext}
which again shows $\kappa_{12}$ is much smaller than $\kappa_{13}$ and $\kappa_{23}$ when $z_{1;ij}$ and $z_{2;ij}$ are small. These expressions look very similar to those given in equations (\ref{eq:kappa12:2}) - (\ref{eq:kappa23:2}) for the two FM case, in fact the only difference is the presence of $J_{0}\left(z_{3;ij}\right)$. But the presence of this new term permits new phenomena because as we vary the wavenumber $q_3$ and/or its amplitude $A_3$ it becomes possible for either $z_{3;13}$ or $z_{3;23}$ to become equal to a zero of the Bessel function $J_0$. The effect will be to either switch off $\kappa_{13}$ or $\kappa_{23}$. If we switch off $\kappa_{13}$ then no transitions between states $1$ and $3$ can occur thus $P_{13}=0$ \emph{even though mode $q_1$ is on resonance}. At this value of $q_3$ the induced transparency has become complete. If we switch off $\kappa_{23}$ then now the effect of the third FM is to switch off the induced transparency effect and so we restore the amplitude of the oscillations of $P_{13}$ to 100\%. We call this effect Restored Opacity. Thus by scanning in the non-resonant mode $q_3$ we can tune the opacity of the system from zero to 100\% even though this mode is nowhere close to being resonant. 
\begin{figure}[t!]
\includegraphics[clip,angle=270,width=\linewidth]{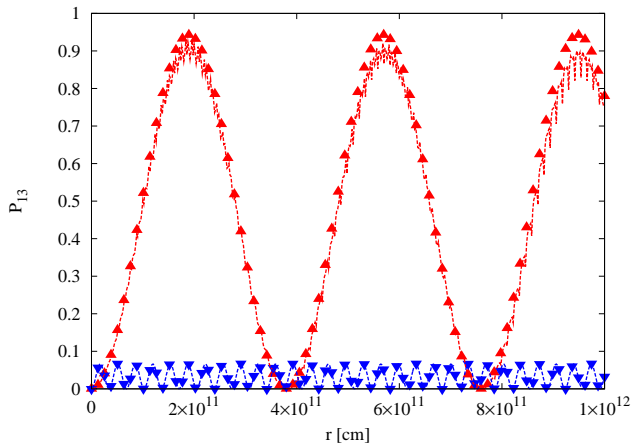}
\caption{The probability $P_{13}$ as a function of $r$ for a three flavor neutrino model. The wavenumbers $q_1$ and $q_2$ are set to $q_1 = \breve{k}_3 - \breve{k}_1$ and $q_2 = \breve{k}_3 - \breve{k}_2$ with amplitudes $A_1=0.02$ and $A_2=0.1$. The third wavenumber is $q_3=5.24\times10^{-10}{\rm cm^{-1}}$. The dashed lines are the numerical solutions, the triangle symbols are the RWA prediction. The blue curve is for the case $A_3 =0$ and produces an example of Induced Transparency. The red curve is for $A_3=0.2$ and produces an example of Restored Opacity. }
\label{fig:ROvsr}
\end{figure}
To test these predictions we solve the for the transition probability $P_{13}$ numerically making no approximation. As for the two FM case, we set the potential $V_{\star}$ to $V_{\star}=6\times 10^{-25}\;{\rm erg}$ and the wavenumbers $q_1$ and $q_2$ are set to $q_1 = \breve{k}_3 - \breve{k}_1$ and $q_2 = \breve{k}_3 - \breve{k}_2$ with amplitudes $A_1=0.02$ and $A_2=0.1$. The third wavenumber $q_3$ is set to $q_3=5.24\times10^{-10}{\rm cm^{-1}}$ and we consider two cases: $A_3 =0$ and $A_3 =0.2$. The comparison between the numerical and RWA solutions is shown in figure (\ref{fig:ROvsr}). In the $A_3 =0$ case we expect induced transparency and indeed the figures shows that is correct with very small amplitude oscillations in $P_{13}$ even though the wavenumber $q_1$ is exactly on resonance between those pair of states. When we switch on the third mode we find $z_{3;23}$ is equal to a root of $J_0$ which means $\kappa_{23}=0$. This should return the amplitude of the oscillations of the transition probability $P_{13}$ back to unity and the figure indicates that does indeed occur: the presence of the third FM with this amplitude and wavenumber leads to a restoration of the opacity. 

\begin{figure}[t!]
\includegraphics[clip,angle=0,width=\linewidth]{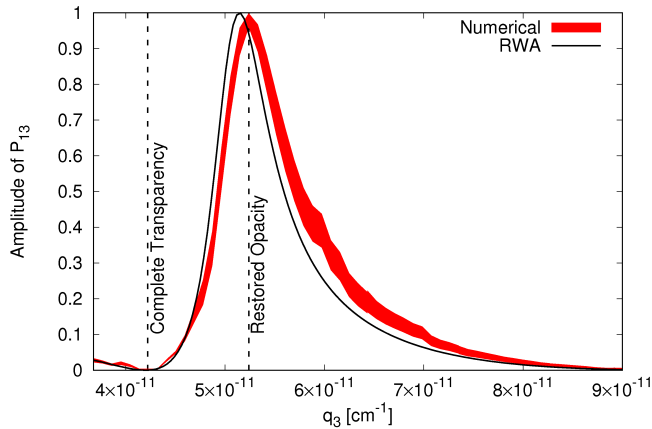}
\caption{The amplitude of $P_{13}$ as a function of the wavenumber $q_3$. The potential $V_{\star}=6\times 10^{-25}\;{\rm erg}$ and the wavenumbers $q_1$ and $q_2$ are set to $q_1 = \breve{k}_3 - \breve{k}_1$ and $q_2 = \breve{k}_3 - \breve{k}_2$ with amplitudes $A_1=0.002$ and $A_2=0.01$. The amplitude of the third FM $q_3$ is $A_3=0.02$. The black solid line is from RWA evaluation and the thick red line is from the numerical solutions with the thickness of the band indicating the width of the fluctuations, an example of which is shown in figure (\ref{fig:ROvsr}). The values of $q_3$ which give Complete Transparency and Restored Opacity are indicated.}
\label{fig:ROvsq3}
\end{figure}

To further illustrate the power of the third FM, in figure (\ref{fig:ROvsq3}) we fix the amplitudes at $A_1=0.002,A_2=0.01$ and $A_3=0.02$, and scan in the wavenumber $q_3$. The purpose of using smaller amplitudes for the FMs is to suppress the fluctuations of the transition probability seen in the numerical results which make it hard to determine the transition amplitude. Note this choice also makes the corresponding value of $q_3$ which cause the Bessel functions to hit their roots smaller than in the example shown in figure (\ref{fig:ROvsr}). From every numerical solution we fit two sinusoids with amplitudes that enclose the oscillations of $P_{13}$ as seen in figure (\ref{fig:ROvsr}). The spread in amplitudes forms the width of the band for the numerical results shown in figure (\ref{fig:ROvsq3}).
The comparison of the theory and numerical solutions in figure (\ref{fig:ROvsq3}) indicate the theory does a very good job of reproducing the numerical results. At $q_3=5.24\times10^{-11}{\rm cm^{-1}}$, $\kappa_{23}$ is zero and therefore opacity is restored. When $q_3=4.22\times10^{-11}{\rm cm^{-1}}$ we find $z_{3;13}$ is a root of $J_0$ which forces $\kappa_{13}$ to be zero and thus we have complete transparency. 


\section{Discussion and Conclusion}
\label{conclusions}

Mathematical tools for calculating the effect of a perturbation upon a quantum system are a valuable resource for both theorist and experimentalist. In this paper we have presented one such tool that is able to take an arbitrary perturbing Hamiltonian that has been decomposed into FMs and predict the evolution upon a multi-flavor neutrino / multi-level quantum system by using the Rotating Wave Approximation. We have placed no restriction on the dimensions of the perturbing Hamiltonian, the number of FMs, the size and structure of the amplitudes nor the wavenumbers. These lack of restrictions make the method applicable to a wide range of problems including those where the Hamiltonian is aperiodic / turbulent (but Fourier-decomposable). 

We illustrated how our method works in practice by using it to analyze how a three flavor neutrino evolves when subject to density fluctuations composed of two and three anharmonic FMs. We found the RWA was able to predict the amplitude and wavenumber of the neutrino transition probabilities between pairs of states to within a few percent and we also discovered the equivalent of electromagnetic induced transparency when the expected maximal oscillations between a given pair of neutrino states could be switched off by the presence of a second resonant FM between another pair of states. When we added a third FM we found the neutrino evolution could be further controlled. At one value for the wavenumber the third FM was able to complete the transparency induced by the second but, at another value, we found the opacity could be restored. The three-flavor effects of induced transparency and restored opacity have obvious application to neutrinos and turbulence and is a area we shall pursue in future work.

\acknowledgments
The authors would like to thanks John Thomas, Bob Golub, Laura Clarke and Jason Bochinski for their helpful comments on this paper. This work was supported at NC State by DOE grants DE-SC0006417 and DE-FG02-10ER41577. 


\end{document}